# Mitigation of crosstalk and noise in multicore fiber on quantum communication

**Ekaterina Ponizovskaya Devine**
*W&Wsens Devices, Los Altos, Davis, CA, 94022*
*Author e-mail address: eponizovskayadevine@ucdavis.edu*

**Abstract:** The influence of crosstalk on quantum communication networks and its mitigation is discussed. It was shown that choosing the parameters for the network that uses the phase stochastic resonance phenomena can increase the signal-to-noise ratio.

1. Introduction

Photonic quantum networks that use entangled photons and photon-enabled quantum entanglement transport and distribution are of great interest [1]. The research studies quantum networks that host multiple services, such as secure communications, distributed quantum computing, and sensing. Recently, several studies addressed the enabling technologies and practical implementation challenges in quantum networking: operation with quantum bits (qubits) and high-dimensional quantum states (qudits), the transmission of qudits over multi-core with the goals to minimization of degradation and decoherence, especially if classical and quantum modes co-propagate. The crosstalk in multicore fiber can produce a false signal creating an error that is simulated in the paper and the ways to mitigate it are discussed.

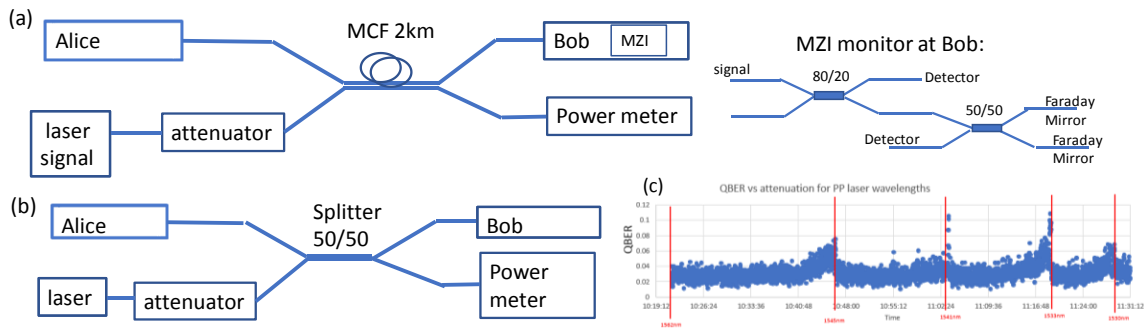

Fig. 1. Quantum communication experimental setup (a)with MCF (b) direct injection, (c) typical QBER recording in time with changing the attenuation of the crosstalk

The simultaneous of classical and quantum signal propagation through the same fiber infrastructure [1] that was the multicore fibers utilized in this hybrid fiber network with multicore fiber (MCF) [2] is discussed. We use quantum communication channel of the quantum key distribution system (QKD) Clavis 3 [3], see Fig.1. We have seen that the stray photons in the fiber were able to increase the error and reduce visibility due to crosstalk Fig1a or a direct weak signal injection Fig1b. The goal is to establish the limits for the classical channel power and the density of the optical channels in the fiber that provide no significant influence on the quantum channel. The quantum channel between Alice and Bob was using the coherent one-way (COW) protocol [3,4]. We use the same setup that was discussed in [5]: at the Bob side there are two Mach-Zehnder interferometers (MZI) are made with two 50/50 couplers and one phase modulator. An emitter at Alice produces qubits states from the computational basis or decoy sequences with the identical time between all consecutive pulses so that the phase relation between those consecutive pulses is kept constant. The receiver analyzes the computational basis and check the phase relation between two consecutive optical pulses. A quantum bit error (QBER) value is measured by counting the probability of having an error in the exchange of qubits of the computational basis; the phase relation check is quantified by measuring the Visibility of interferences occurring in the second basis analyzer. Based on both values, QBER and visibility, it is possible the estimate if it is possible to extract secret keys form the qubits exchanged between the emitter. The results of the experiment shows that the signal in MCF can influence on the QBER and Visibility that interrupts the key generation. The study the ways to mitigate the crosstalk influence. One way is using the fiber that reduces the crosstalk [4] and another is to study the conditions for the stochastic resonance that can increase the signal-to-noise ratio [5].

## 2. Crosstalk reduction

There are Using the Beam propagation method (BPM) [6] we estimate the crosstalk between the fiber cores (Fig.2) and compared it to the experiment. We had studied the techniques to reduce the crosstalk. First, the BPM simulations were validated with experimental measurements of the crosstalk in the fiber without trenches (Fig.2a). The intensity in dBm for the 7 core fiber is shown in Fig.2d and e for the signal in the central core and in one of the peripheral cores.

Fig. 2. BPM crosstalk simulation: MCF cross section(a) BPM simulated intensity in MCF 4-core cross section (b) 7-core MCF (c) intensity in MCF in dBm for signal in different cores (d) and (e).

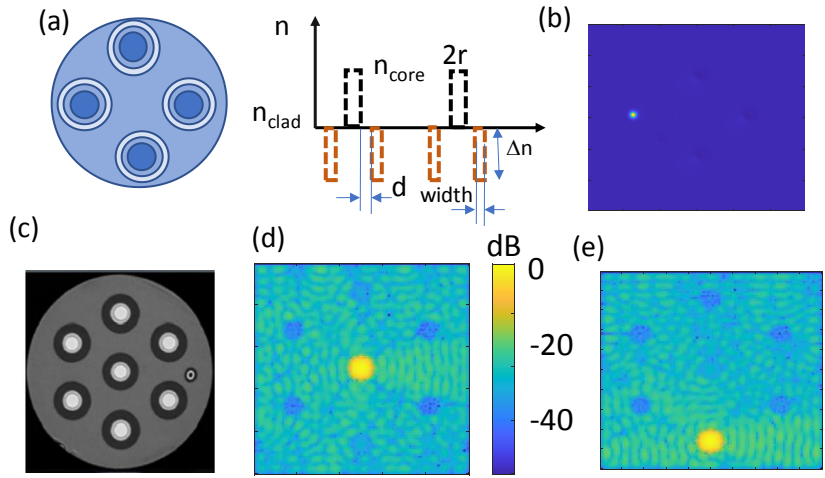

The simulations are done for the parameters typical for the single mode MCF, where the cladding diameter is 125μm, a square lattice with a period of 50μm in the 4-core fiber. The core diameter is 7 μm [4]. We study the single-mode propagation at 1550nm wavelength through these fibers as well as the multiple-mode propagation with crosstalk. The experimental setup consisted of the laser with wavelength 1550nm and power varied from 0 to 20mW, on the other end of 1 m and 2 km fibers the signal was measured in the all for cores. The results were consistent with the simulations. The higher core density results in higher crosstalk.

Table 1. Crosstalk BPM simulations for MFC with and without trenches

| Width (micron) | Crosstalk (4-core) (dB) (dn 0.005/0.01) |
|---|---|
| 0 | -50 |
| 1 | -52/-59 |
| 3 | -55/-75 |
| 6 | -62/-79 |

We also study the limits on the distance between the cores and the crosstalk reduction due to "trenches" [6] with a refractive index lower than in the cladding around cores, see Fig.1c. The fiber was modeled with BPM. The results of BPM are sown in Table 1 for various trenches width and index difference.

### 3. QKD modeling and experimental validation

The quantum channel (QC) between Alice and Bob of a QKD system Clavis-3 is used to study the communication between through a 2km multicore fiber as it is shown in Fig.1 (a). One of the cores is used for QC and other cores transmit a signal from a laser that goes through the attenuator and is measured by the power meter on the other end of the MCF fiber. The experiment was done with NKT Photonics laser, Pure Photonics (PP) laser in whisper and normal modes, and RIO laser. Those lasers are all low noise, however, the PP laser had a slightly wider noise bandwidth (about 10-15KHz) while RIO and NKT noise bandwidth was only 1-2kHz.

A power meter was used on the free end of the 50/50 splitter to control the laser power. QBER and Visibility were measured on the Clavis-3 system. Usually, QBER is about 2%. It sharply increased after we start transmitting the laser signal and then stabilized at 4% (Fig.1c). PP laser had a variable wavelength in a range from 1530 to 1560 nm. The QKD system transmits its signal at 1550nm wavelength and has a bandpass filter at about 1550nm. We changed the wavelengths and have seen the response in QBER and Visibility. Each frequency stays for 5 min then the laser is disabled, the frequency switched to the next, and enables again. The Key generation never was interrupted although we noticed an increase in QBER. We simplified the experiment by producing the direct injection of an external laser in QC through the attenuator to reduce the signal to values similar to the crosstalk values.

The model of the QKD quantum channel with crosstalk from the multiple external signals was developed [5] and tested for the direct injection experiment (Fig. 1b) with small noise.
The visibility with several crosstalk signals from multiple cores *i* can be defined as

$$V = (D_2 - D_1 + \alpha \sum_i P_{li}(cos(\theta_i + 4\pi f_{ni}\Delta T)))/(D_2 + D_1 + \alpha \sum_i P_{li}) \quad (1)$$

where $D_1$ and $D_2$ are the probability of the clicks if we have only a QKD signal without crosstalk, the delay time of MZI is $\Delta T$ = 50ps, a phase shift of $\vartheta_\omega = \omega\Delta T/2$, wavelength difference between signals is $\Delta\omega$. The phase noise of the signal is characterized as a frequency *f* distribution with a mean value $<f_n> = 0$ and some mean square $\sigma^2 = \langle f_n^2 \rangle$ that is the width of the linewidth of the external signal. The frequency mismatch can be characterized by $V_\omega = cos^2\vartheta_\omega - sin^2\vartheta_\omega$ that varies from -1 to 1, $P_l$ is the external laser power and $\alpha$ is a normalization to the amplitude of the quantum signal. The inverse threshold of crosstalk signal, when communication is lost, is represented by the parameter $(\alpha\sum P_{li})^{-1}$ in Eq.1, with a Visibility threshold of 80%. The key can't be generated below the threshold

Fig. 3: The visibility recorded experimentally and theoretically with Eq(1).

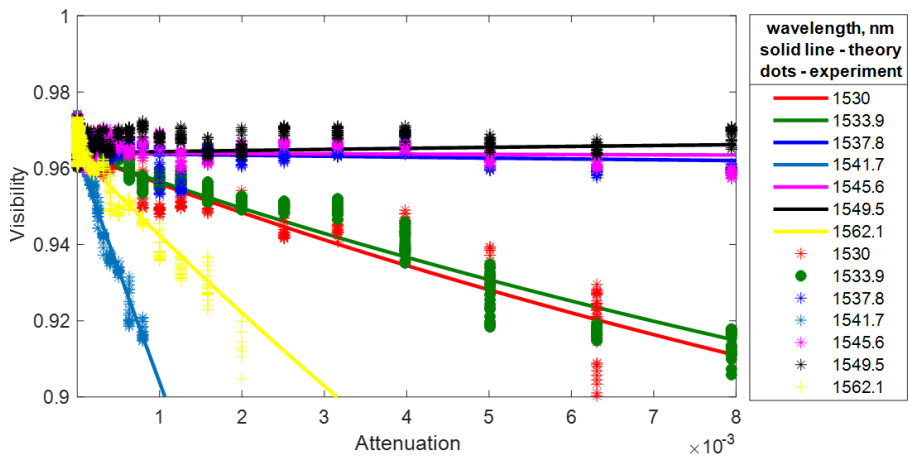

value.

The experimental data for the direct injection are consistent with the crosstalk data that showed the effect starting at about the power -40-50dB. The key distribution can be disrupted at the higher power. However, the threshold was 11-15dB lower for RIO and NKT lasers as well as for certain wavelengths at PP lases. It was shown that it is the result of the wavelength of the external signal and the difference in noise was too small to cause the results. The data collection for QBER and Visibility vs attenuation for PP laser whisper mode is shown in Fig.3 in log scale where it is close to a linear dependence of the attenuation. The solid lines in Fig.2c show the theoretical visibility from Eq.1. compared with dots that represent the experimental data. The values $D_1$ and $D_2$ are taken from the Visibility of the QKD system without an external signal. The parameter $α$ was fitted and was found to be 15dB. The same value of the fitting parameters was valid for other measurements. The experimental data fit the model. Wavelength shift as low as $\Delta\lambda = 0.3nm$ changes $V_\omega$ from plus to minus it can switch the threshold from -40dB (light blue curve) to less than -10dB (black, magenta, dark blue curves). The effect of the narrow filter at QKD Bob entrance showed as the increase of parameter $α$ for the wavelengths near 1550nm. The phase noise was too small to affect the results significantly.

## 4. Phase stochastic resonance

Stochastic resonance is a phenomenon that still inspires wide interest both as a scientific phenomenon and for its applications that occurs in non-linear bi-stable systems [7] as well as in a linear system with multiple external signals [8] and is extremely important for quantum network design. The stochastic resonance in quantum communications networks wasn't been significantly studied yet. It can help to achieving higher signal-to noise ratio (SNR) that is extremely important for quantum network design. In the real system the phase noise of the signals in the neighboring cores can be of several THz for femtosecond pulses. We did the simulation for various wavelengths and noise. The SNR was defined as a threshold crosstalk when communication is lost.

Fig. 4. Phase stochastic resonance with crosstalk from 3 cores with few sets of carrier wavelengths and phase noise (a), the position of PSR depending on the wavelength shift $V_w$ for crosstalk from 2 other cores (b)

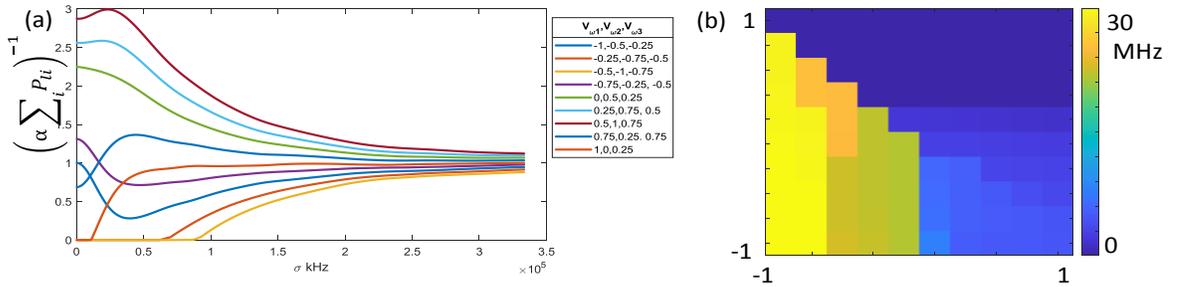

The threshold value of the ration of quantum signal to crosstalk amplitude $(\alpha P_l)^{-1}$ in Eq (1), that represents the parameter with a Visibility drops to 80%. The key can't be generated below the threshold value stopes and the connection between Alice and Bob is llost. The results of the simulation for different frequency mismatch is shown in Fig 4a for a cross-talk from 3 cores with different frequencies and the same phase noise. As we can see some frequency mismatch combinations produce the lower or higher threshold (normalized to the value at infinite noise) at phase noise with a maximum in a certain range that indicates the phase stochastic resonance (PSR). Fig.4b shows the position of the PSR in the space of $V_w$ for crosstalk from 2 cores. The harmonic term in Eq. 1 was averaged using Monte-Carlo method.

Fig. 5. SNR vs phase noise bandwidth for several combinations of wavelength. The SNR is increasing with $V_{w2}$ until it creates a region of noise where the crosstalk doesn't influence the quantum channel (yellow curve).

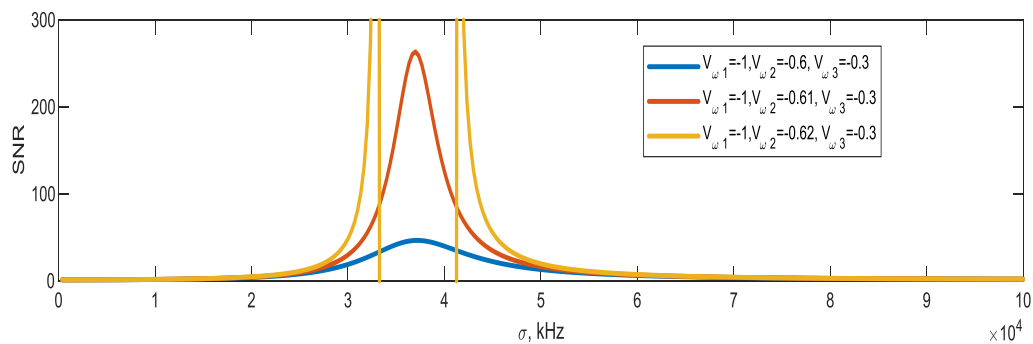

As we can see there are the parameters combination with zero threshold that means the crosstalk doesn't affect the quantum channel. The SNR curve with PSR depending on the wavelength of the crosstalk is presented in Fig.5. At certain wavelength the SNR start increasing and for some wavelength there is a range of phase noise where the crosstalk does not affect the communication.

## 5. Conclusions

Our study shows, that SNR at a given wavelength of the external signals has a maximum for a certain range of noise. The results can be interesting for the design of quantum communication systems and put restrictions on the signal parameters for the optimal configuration. The BPM simulation shows that tranches can reduce the crosstalk to up to -80dB instead of -50dB that could mitigate the problem and increase SNR. It was shown the effect of the phase stochastic resonance for certain parameters of the crosstalk.